\documentstyle [12pt] {article} 

\newcommand{\sect}[1]{ \section{#1} } 
 
\newcommand{\req}[1]{(\ref{#1})} 

\newcommand{\ve}{\left( \begin{array}{r}} 
\newcommand{\ev}{\end{array} \right)} 
\newcommand{\ar}{\left( \begin{array}{rr}} 
\newcommand{\ra}{\end{array} \right)} 
\newcommand{\arr}{\left( \begin{array}{rrrr}} 
\newcommand{\arrr}{\left( \begin{array}{rrrrrr}} 

\newcommand{\eqr}{\begin{eqnarray}}
\newcommand{\rqe}{\end{eqnarray}} 
\newcommand{\eq}{\begin{equation}} 
\newcommand{\qe}{\end{equation}}

\newcommand{\half}{\frac{1}{2}}

\newcommand{\quart}{\frac{1}{4}}

\newcommand{\ket}[1]{\mbox{$ \mid #1 >$}}

\newcommand{\pr}{p_R}
\newcommand{\pl}{p_L}
\newcommand{\prt}{\tilde{p}_R}
\newcommand{\plt}{\tilde{p}_L}
\newcommand{\mhat}{\hat{m}}
\newcommand{\nhat}{\hat{n}}
\newcommand{\lhat}{\hat{l}}
\newcommand{\rhat}{\hat{r}}
\newcommand{\that}{\hat{t}}
\newcommand{\what}{\hat{w}}

\newcommand{\ig}{g^{-1}}
\newcommand{\ih}{h^{-1}}
\newcommand{\ic}{c^{-1}}

\begin{document}

\begin{flushright}
hep-th/9606145
\end{flushright}

\vspace{2cm}

\begin{center}
{\bf\Large 
Generalized string compactifications with spontaneously broken
supersymmetry
} 

\vspace{1.5cm}

{\small
{\bf Hans-Peter Nilles}\\
\vspace{0.2cm}
        Physik Department\\
        Technische Universit\"at M\"unchen\\  
        D--85748 Garching, Germany
\\
and
\\ 
Max-Planck Institut f\"ur Physik\\
F\"ohringer Ring\\
D-80805 M\"unchen, Germany.\\

\vspace{0.8cm}

{\bf Micha\l\ Spali\'nski}\\
\vspace{0.2cm}
Institute of Theoretical Physics\\
Warsaw University\\
Warsaw, Poland.
}
\end{center}

\vspace{1.2cm}

\begin{abstract} 

The Narain lattice construction of string compactifications is
generalized to include spontaneously broken supersymmetry. Consistency 
conditions from modular invariance and Lorentz symmetry are solved in
full generality. This framework incorporates models where
supersymmetry breaking is inversely proportional to the radii of
compact dimensions. The enhanced lattice description, however, might
allow for models with a different geometrical or even non-geometrical 
interpretation. 

\end{abstract}

\thispagestyle{empty}

\newpage

\setcounter{page}{1}

\sect{Introduction}

The problem of supersymmetry breaking in string theory is one of the
basic problems the theory faces if it is to be relevant for elementary
particle physics. One possibility, which has to some extent been 
explored in the past \cite{rohm,fkpz,a}, is that supersymmetry may be
broken spontaneously at tree level in string theory. This possibility
is actually rather special, since it is known \cite{dix} that
vacua with broken supersymmetry cannot be continuously connected to
ones where supersymmetry is unbroken. The only possibility which the
theorem leaves open is that the broken and unbroken vacua are
connected, but one cannot be reached from the other by a sequence of
infinitesimal transformations. This is realized,
for example, when the supersymmetry breaking order parameter vanishes
as one of the compactification radii goes to infinity: only in this
decompactification limit is supersymmetry restored.

This paper presents a systematic method of constructing a large class
of N=4 models with spontaneously broken supersymmetry, which can also
be twisted to reduce the number of supersymmetries. The approach
followed here is a direct generalization of the work of Narain et al.
\cite{nara,nsw} in the spirit of \cite{grv,ms2,erspa}. In this
approach, 
all toroidal compactifications are described by specifying a family of
Narain lattices parameterized by a set of moduli. These can either
appear in the basis vectors of the lattice, or the basis can be
constant over moduli space while the moduli enter as components of
various background fields. We will make use of both these pictures.
The virtue of the Narain formulation of toroidal string vacua is that
the moduli dependence of the theory is explicit. A given model is not
locked into some special point in moduli space, as is the case for
covariant lattice constructions \cite{lls}, or the free fermionic
construction \cite{klt,abk}. 

The Narain lattice (for the case of a $4$
dimensional compactification) has dimension (6,22). The worldsheet NSR
fermions play no role in the compactification process -- that sector
is the same in the compact theory as in the non--compact case. The
basic idea put forward here is to bosonize the NSR fermions and treat
them on the same footing as the other compact coordinates. These
additional bosonic degrees of freedom will be referred to as ``NSR
bosons'' in the sequel. In particular, the construction outlined below
allows backgrounds which mix these additional directions in the
lattice with the ones present already in the work of Narain, Sarmadi
and Witten. It is this new class of backgrounds which makes
spontaneous supersymmetry breaking possible.

The key issues are whether modular invariance and Lorentz covariance can be
maintained in the presence of the new backgrounds. It is straightforward to
check that one-loop modular invariance holds\footnote{In this communication
we do not address the issue of higher loop modular invariance.} for a
nontrivial class of backgrounds. These new backgrounds bear a formal
analogy to the Wilson lines; however rather than correlating the space
compactification lattice with the $E_8\times E_8$ lattice, they connect it
with the part corresponding to the NSR bosons. In what follows, the new
backgrounds will be referred to as generalized Wilson lines. In accordance
with general theorems these extra background parameters will be seen to be
quantized in order to preserve Lorentz covariance.

The construction presented here gives theories with N=4 supersymmety.
Some of these supersymmetries are spontaneously broken for an
appropriate choice of the backgrounds (or equivalently, of the Narain
lattice). It is natural to look at orbifolds of these models which
would give 
spontaneously broken N=1 supersymmetric theories.\footnote{The class of
  theories obtained in this way includes the examples of references
  \cite{fkpz,a}.} This should be straightforward to achieve using, for
example, the approach of \cite{ms2,erspa,erk}. The formulation of the
problem 
presented here is set up so as to carry out this step. The results of
that study will be presented in a more detailed communication in the
future.

The constructions of string compactifications with spontaneously broken
supersymmetry which exist in the literature were carried out via so
called coordinate dependent compactifications \cite{fkpz} from 5 to 4
dimensions, first in the context of fermionic string models, and more
recently also for Narain models \cite{ben}. These constructions
involve specially chosen U(1) currents -- these currents correspond to
particular directions in the extended Narain lattices introduced in
this paper.

The construction given here works for compactifications down to any
dimension $2<d<10$. The discussion is presented for compactification down
to $d=4$ to be specific.  Apart from generalizing known examples and giving
a systematic means of studying models with spontaneously broken symmetry in
a familiar framework, the construction presented here could play a role in
finding pairs of dual string theories with a reduced number of
supersymmetries \cite{vafwit,vafsen}. Indeed, it is natural to consider
strings compactified to low dimensions with reduced supersymmetry: such
compactifications would then lead to higher dimensional theories with
higher numbers of supersymmetries in various ``decompactification''
limits. It is also interesting to see how the models described here fit in
with the general considerations of string spectra \cite{d1}. In particular
it would be relevant for phenomenology to check whether the supertrace of
the mass square matrix indeed vanishes in these models \cite{d2}.

The paper is structured as follows: section $2$ describes a fully
bosonized incarnation of the heterotic string compactified on a Narain
lattice in a form suitable for further developments. In section $3$
the generalized Wilson lines are introduced and consistency with
one-loop modular 
invariance and Lorentz symmetry is discussed. Section $4$ presents
some general considerations of the spectrum and shows how
supersymmetry breaking can take place in these models.

\sect{Bosonization}

This section describes how the Narain compactified heterotic string
can be formulated in purely bosonic terms. The fact that this can be
done is well known -- the reason for reviewing the subject here is to
extend the framework of \cite{ms2} to include bosonized fermions, so
that the generalization discussed in the following section can be
easily stated.

The heterotic string in the light front gauge can be described purely
in terms of the following bosonic fields:

\begin{enumerate}
\item Non-compact (transverse) spacetime coordinates $X^a$, $a=1,2$.
\item Left and right--moving bosons corresponding to ``internal''
  degrees of freedom $X^i$, $\bar{X}^i$, $i=1..6$.
\item Left--moving bosons $\bar{Y}^I$, $I=1..16$, compactified on the
  $E_8\times E_8$ lattice.
\item Right--moving bosons $H^A$, $A=1..4$, which bosonize the
  world-sheet NSR fermions (the ``NSR bosons'').
\end{enumerate}

The zero-modes of the compact fields lie on a generalized Narain
lattice of dimension $(10,22)$. The possibility of writing the
partition function for the heterotic string in the light--front gauge
in terms of these degrees of freedom has been noted in
\cite{klt,lls}. 

For the moment only the metric moduli of the $(6,6)$ compactification
lattice will be considered. The remaining moduli of the Narain lattice
-- the antisymmetric 
tensor field and the Wilson lines will be incorporated later. 

The spectrum of the theory follows from the Virasoro
generators, which can be written as \footnote{All formulae assume 
$\alpha^\prime = 1$.} 
\eqr
L_0 &=& \quart p^2 + L_0^\prime + N -\half , \\
\bar{L}_0 &=& \quart p^2 + \bar{L}_0^\prime + \bar{N} - 1 ,
\rqe 
where
$p$ is the spacetime 4-momentum, $N, \bar{N}$ are integer valued
oscillator number operators, and the (compact) zero-mode
contributions read (in matrix notation) 
\eqr
L_0^\prime &=& \quart \pr^T g^{-1} \pr + \half \prt^T \ih \prt , \\
\bar{L}_0^\prime &=& \quart \pl^T g^{-1} \pl + \half \plt^T \ic \plt . 
\rqe 
Here 
\eqr
\pr &=& \mhat - g \nhat ,\\
\pl &=& \mhat + g \nhat ,
\rqe 
and 
\eqr
\prt &=& \rhat + \that , \\
\plt &=& \lhat ,
\rqe 
are the momenta in the lattice basis. The hatted
quantities have integer entries apart from the vector $\that$, which
is given by 
\eq 
\that = (-\half, -\half, -\half, 1) . 
\qe 
In the following
it will sometimes be convenient to use the notation
\eq
\what \equiv \rhat + \that ,
\qe 
which is also not integer.

The shift vector $\that$ appears because of the spacetime fermions present
in the spectrum of the heterotic string. In a canonical basis,
described later on, the shift by $\that$ is responsible for
transforming the $o$ and $c$ cosets of the $D_4$ lattice to the $v$
and $s$ cosets 
actually appearing in the physical partition function. The vectors
$\mhat, \nhat$ are the momenta and windings of the $6$ compactified
coordinates, $\lhat$ are the momenta of the $Y$ bosons, while $\rhat$
denote the momenta of the chiral bosons $H$ corresponding to the NSR
fermions. These are all (column) vectors of dimensions $6,6,16,4$
respectively.

The matrices $g, h, c$ are the metrics of various parts of the full
lattice. The matrix $g$ is a real symmetric matrix containing the
metric moduli of the $(6,6)$ compactification lattice of the internal
compact coordinates. The matrices $c,h$ are fixed, integer matrices
whose form is required by modular invariance: $c$ is the $E_8\times
E_8$ lattice Cartan matrix, while $\ih$ is given by 
\eq 
\ih = \arrr
2&1&1&1\\1&2&1&1\\1&1&2&1\\1&1&1&1 \ra .
\qe
The form of this metric can
be found, for example, by studying the Poisson resummed partition
function in the standard formulation.

It is often convenient \cite{ms2} to use the quantities $H$ and $P$
defined by 
\eqr
H&=&L^\prime_0 + \bar{L^\prime}_0 , \\
P&=&L^\prime_0 - \bar{L^\prime}_0 .
\rqe 
Note that only the zero-mode
contributions enter here -- the oscillators and normal ordering
constants do not appear in this definition. One finds 
\eqr
H &=& \half (u+s)^T \chi (u+s) ,\\
P &=& - \half (u+s)^T \eta (u+s) ,
\rqe 
where 
\eq u=\ar \mhat \\ \nhat \\ \lhat \\ \rhat \ra , 
\qe 
is a vector of $32$ integers grouping all
the zero-mode quantum numbers of the various string states, while 
\eq 
s = \ar 0 \\ 0 \\ 0 \\ \that \ra .
\qe 
The matrix $\eta$ is the constant Narain lattice
metric\footnote{Note that the extended Narain lattice is no longer even.}: 
\eq 
\eta=\arrr
0&1&0& 0\\
1&0&0& 0\\
0&0&\ic& 0\\
0&0&0&-\ih 
\ra ,
\qe 
while $\chi$ contains all the modulus dependence: 
\eq
\label{trivial}
\chi=\arrr
g^{-1}&0&0&0\\
0&g&0&0\\
0&0&\ic&0\\
0&0&0&\ih 
\ra .
\qe 
When other moduli are given non-vanishing values, the
matrix $\eta$ retains its (constant) form -- the moduli always appear
in $\chi$ only. Thus the issue of introducing new backgrounds can be
formulated as finding forms of the background matrix $\chi$ which
satisfy the requirements of modular invariance.

To check modular invariance one has to study the behaviour of the
partition function of the theory under transformations of the
worldsheet modular parameter $\tau$.  Using the objects given above, 
the moduli dependent zero-mode contribution to the one-loop partition
function can be written 
in the form 
\eq 
\label{pf}
P(\tau,\bar{\tau}) = \sum_u \exp\{-\pi (u+s)^T
M(\tau,\bar{\tau}) (u+s) + 2\pi i u^T\eta s\} ,
\qe 
where 
\eq
M(\tau,\bar{\tau}) = i Re(\tau) \eta + Im(\tau) \chi . 
\label{mdef}
\qe 
The term linear in $u$ which appears in the exponential in \req{pf} is
responsible for the GSO projection in the spectrum. 
It is
straightforward to check that this leads to the known result for the
toroidally compactified heterotic string one-loop partition function,
expressed in terms of Jacobi 
theta functions, the $E_8\times E_8$ lattice sum and a soliton sum
over the $(6,6)$ compactification lattice.

Modular invariance of these theories can be demonstrated using the
following remarkable properties of the matrix $M$: 
\eqr
M(\tau+1, \bar{\tau}+1) &=& M(\tau,\bar{\tau}) + i\eta ,\label{crux1}\\
M(-\frac{1}{\tau}, -\frac{1}{\bar{\tau}}) &=& \eta
M^{-1}(\tau,\bar{\tau}) \eta .
\label{crux2}
\rqe 
The first of these properties is true for any background, as is clear
from the definition \req{mdef}. The second property, however, places
stringent restrictions on the possible background matrices $\chi$. 

When nontrivial antisymmetric tensor and Wilson line backgrounds are 
introduced in the correct way, these transformation properties of $M$
remain unaltered. 

The Narain models of \cite{nsw} are expressed in this language by the
following form of the background field matrix \cite{ms2,erspa} : 
\eq 
\chi
= \arrr
\ig & -\ig b^\prime & -\ig a^T \ic & 0 \\
{- b^\prime}^T \ig & (g + {b^\prime}^T)\ig (g+b^\prime) &
(g+{b^\prime}^T)\ig a^T \ic & 0\\
- \ic a\ig & \ic a\ig (g+b^\prime) & \ic+\ic a \ig a^T \ic & 0\\
0&0&0&\ih \ra ,
\qe 
where 
\eqr
b^\prime &\equiv& b+\half a^2 , \\
a^2 &\equiv& a^T \ic a . 
\rqe 
Here $b$ is an antisymmetric matrix ($6$ by
$6$) containing the ``axionic'' moduli and $a$ is the matrix ($16$ by
$6$) containing the Wilson line moduli. One may check directly that
\req{crux2} is fulfilled.

\sect{Generalized Narain Models}

This section presents the main technical development of this paper.
We consider background field
matrices $\chi$ which involve lattice directions corresponding to the
zero-modes of the NSR bosons. It is clear that the
appearance of such backgrounds will generically lead to the breaking
of worldsheet superconformal invariance and consequently to the
spontaneous breaking of spacetime supersymmetry. It will later be shown
that in order that these backgrounds do not lead to a model with
broken Lorentz symmetry, the new background parameters have to be
quantized.

The new moduli are grouped into a $4$ by $6$ matrix denoted by $x$. 
To simplify the formulae somewhat, the ordinary Wilson lines have been
set to naught, since in any case they will not be discussed further
here. The modified background matrix, including the generalized
Wilson lines $x$, reads:
\eq 
\label{fullchi}
\chi = \arrr
\ig & - \ig b^\prime & 0 & \ig x^T \ih \\
- {b^\prime}^T \ig & (g - {b^\prime}^T)\ig
(g-b^\prime) & 0 & (g -{b^\prime}^T)\ig x^T \ih \\
0 & 0 & \ic & 0 \\
\ih x \ig & \ih x \ig (g -b^\prime) & 0 & \ih + \ih x \ig x^T \ih \ra .
\qe 
Here
\eq
b^\prime \equiv b - \half x^2 ,
\qe
and 
\eq
x^2 \equiv x^T \ih x . 
\qe
It is straightforward to verify that the new background fields $x$
enter in such a way that the 
crucial property \req{crux2} remains intact. This way
one ensures that the resulting partition function is modular
invariant. 

To make the formulae more readable it is convenient at this stage to
introduce a (moduli dependent) canonical basis defined by the condition
that in that basis the metrics $g, h, c$ should be unit matrices. When
referring to this basis, the vectors $\mhat,\nhat,\lhat,\what$ will
be written without the hats, while the $32$ dimensional objects like $u$ or
$\chi$ will be written with a tilde over them. The background fields, and
the left and right momenta 
hitherto written in lowercase letters, will be written in capitals when
expressed in the canonical basis. The basis transformation matrices satisfy 
\eq
g = e_g^T e_g, \hspace{0.6cm}h=e_h^T e_h, \hspace{0.6cm}c=e_c^T e_c,
\qe
so that
\eqr
m &=& e_g^\star \mhat , \nonumber \\
n &=& e_g \nhat , \nonumber \\
w &=& e_h^\star \what , \nonumber \\
l &=& e_c^\star \lhat 
\rqe
where $A^\star$ denotes $(A^T)^{-1}$. 

For future reference let us note, that
the basis transformation matrix relating the lattice basis and the
canonical basis in the NSR block is given by
\eq
e_h^\star = \arrr -1& -1& -1& -1/2 \\ 1 &0 &0 &1/2 \\ 0 &1 &0 &1/2 \\ 0 &0
&1 &1/2 
\ra .
\qe
The generalized Wilson line is transformed as
\eq
X = e_h^\star x e_g^{-1} ,
\qe
so in the canonical basis this depends on the metric moduli. 

For the present purpose it will suffice to keep just the generalized
Wilson lines, setting the
antisymmetric background to naught. Using the notation 
\eq
X^2\equiv X^T X ,
\qe 
the background matrix in this basis is 
\eq 
\tilde{\chi}=\arrr
1&\half X^2 & 0 & X^T\\
\half X^2 & (1+\half X^2)^2 & 0 & (1+\half X^2) X^T\\
0&0&1&0\\
X& X(1+\half X^2) &0&1+XX^T \ra .
\qe
The various lattice metrics do not appear explicitly here, but the
consequence is that $X$ is now moduli dependent. 

While one-loop modular invariance is guaranteed by construction for all
backgrounds of this kind, the requirement of Lorentz covariance of the
theory introduces further restrictions.  The preservation of Lorentz
symmetry in the light-front gauge requires that the supercurrent be well
defined\footnote{One of us (M.S.) would like to thank I. Antoniadis for an
illuminating discussion on this subject.}. This means that under
$\sigma\rightarrow\sigma+2\pi$ the supercurrent should either be periodic
(in the 
Neveu--Schwarz sector) or anti-periodic (in the Ramond sector). To check
what this implies for the generalized Wilson lines, note that the
supercurrent has the form 
\eq 
T_F = \sum_{A=1}^{4} \exp (iH^A) \partial {\bf X}^A .
\qe 
The ${\bf X}^A$ are complex linear combinations of the
original $X$'s. This follows from the usual form of the supercurrent in the 
fermionic language via the bosonization relations
\eq
\psi^A = \exp (iH^A) ,
\qe  
where $\psi^A$ are the (complex) NSR fermions. The supercurrent boundary
condition puts constraints on the allowed spectrum of zero-modes 
of the NSR bosons, that is, on the allowed eigenvalues of $\prt$. From the
explicit form 
of the background matrix $\chi$ \req{fullchi} it follows that 
\eq 
\tilde{p}_R = \what + x \nhat .
\qe 
The supercurrent condition requires that in the
canonical basis these vectors should have either all integer components, or
all half-integer components. Explicitly, 
\eq
\tilde{P}_R = w + e_h^\star x \nhat .
\qe
The states in the Hilbert space for which
$\tilde{P}_R$ is integral belong to the Neveu-Schwarz sector, and the 
ones for which it is half integral belong to the Ramond sector.
This quantization condition on the allowed spectrum of $\tilde{P}_R$
translates 
immediately to a quantization condition on the generalized Wilson lines
$x$. 

Therefore the generalized Wilson lines do not describe
continuous deformations of the extended Narain lattice, but rather
parameterize a discrete family of extended Narain lattices. 

\sect{The spectrum}

To see the impact of the new background on the spectrum of the model
it is useful first to recall what the spectrum looks like in the case
of a purely metric background.

In full generality the spectrum is given by 
\eqr
\half M^2 &=& H + N + \bar{N} - \frac{3}{2} , \nonumber \\
0 &=& P + N - \bar{N} + \half  ,
\rqe 
where $H$ contains all the
dependence on the backgrounds via $\chi$.  Thus we have \eqr
\label{fullspec}
\half M^2 &=& \half (u+s)^T \chi (u+s) + N + \bar{N} - \frac{3}{2} ,
\nonumber \\
0 &=& -\half (u+s)^T \eta (u+s) + N - \bar{N} + \half .
\rqe 
Inserting
the purely metric background given by \req{trivial} and going to the
canonical basis one finds 
\eqr
\half M^2 &=& \half(m^2 + n^2 + l^2 + w^2) + N + \bar{N} -
\frac{3}{2} ,\nonumber \\
0 &=& - mn + \half (w^2 - l^2 + 1) + N - \bar{N} .
\label{spec}
\rqe
Recall here that 
\eq 
w=r+t 
\qe 
is a vector belonging to the $v$
or $s$ coset of the $D_4$ lattice.  This gives the known
massless spectrum of toroidal compactifications, with an $N=4$
supersymmetric structure.

Indeed, for massless states with $m, n = 0$ the formulae \req{spec}
are equivalent to 
\eqr
N &=& \half (1 - w^2) , \\
\bar{N} &=& 1 - \half l^2 .
\rqe 
On the right, massless states have
$N=0$ and $w^2=1$. The latter condition has $16$ possible solutions:
\eq
w=(\pm 1, 0, 0, 0) 
\qe
and permutations (8 vectors in all), and
\eq
w= (\pm \half, \pm \half, \pm \half, \pm\half) 
\qe 
with an even number of minus signs (another 8 vectors).
These solutions express
the supersymmetry of the spectrum: the states come in multiples of
$16$ ($8$ bosonic degrees of freedom, $8$ fermionic).

On the left one has either $\bar{N}=0$ or $\bar{N}=1$, with $l^2=2$ or
$l^2=0$ respectively.

To summarize, the spectrum for $x=0$ is $N=4$ supersymmetric and
consists of the following states:
\begin{itemize}
\item The graviton multiplet ($2\times 16$ states): 
\eq 
\bar{X}^a_{-1}  e^{iwH} \ket{0} .
\qe
\item The gauge sector ($(480 + 16)\times 16$ states) 
\eqr
  e^{il\bar{Y}} &e^{iwH}& \ket{0} ,\\
  \bar{Y}^I_{-1} &e^{iwH}& \ket{0} .
\rqe
\item The matter sector ($6\times 16$ states): 
\eq 
\bar{X}^i_{-1}  e^{iwH} \ket{0} .   
\qe
\end{itemize}
($\bar{X}_n$ and $\bar{Y}_n$ denote the mode operators of the string
coordinates 
and the vectors $w$ appearing here span the $16$ possibilities given 
earlier). 

The full spectrum for non-vanishing background $x$ can be obtained by
substituting the form of the 
background matrix given by \req{fullchi} into the general formula
\req{fullspec}. Setting the antisymmetric background to naught for
simplicity, the resulting equations can be written in the form
\eqr 
\half M^2 &=& \half (m + X^T w + (1+\half X^2) n)^2 + l^2 + 2
\bar{N} - 2 , \\
0 &=& - m n - \half (l^2 - w^2) + N - \bar{N} + \half .
\rqe 
To show explicitly where the metric $g$ enters it is better to rewrite this
in the
lattice basis:
\eqr 
\half M^2 &=& \half (\mhat + x^T \ih \what
+ (g+\half x^2) \nhat)^T \ig (\mhat + x^T \ih \what + (g+\half x^2)
\nhat) \nonumber \\ 
&+& \lhat^T \ic \lhat + 2 \bar{N} -
2 , \\
0 &=& - \mhat^T \nhat - \half (\lhat^T \ic \lhat - \what^T \ih \what) + N -
\bar{N} + \half .
\rqe 
This makes it possible to study various decompactification
limits where supersymmetries are restored.

The above formulae demonstrate that depending on $x$ some of the states
which are massless at $x=0$ become massive, with masses proportional to the
inverse radius of the compact manifold. Thus, for an appropriate
choice of $x$, one gets a behaviour reminiscent of Scherk -- Schwarz type
symmetry breaking in field theory\cite{sch}. In particular, 
one sees that the gravitinos can acquire masses. Thus, the framework put
forward here is geared toward analyzing various patterns of spontaneous
supersymmetry breaking. This becomes particularly interesting in the case
when orbifold twists are introduced.

\sect{Conclusions}

We have formulated a class of models with spontaneously broken
supersymmetry. This was done in the framework of Narain--like models
on a Lorentzian $(14-d,26-d)$ lattice, where $d$ is the number of
non--compact dimensions.  For compactifications down to $d=4$ this
leads to a $(10,22)$ lattice instead of the usual $(6,22)$. Such an
extended 
lattice has additional ``moduli'', the generalized Wilson lines. For
consistency of the theory these have to be quantized, so the actual
number of continuous moduli is the same as in the Narain case. One has
therefore an infinite, discrete family of extended Narain lattices,
each deformed in terms of the standard moduli.

The new backgrounds lead to (symmetry breaking) shifts in the spectrum
which are proportional to $g^{-1}$. Thus, these effects will disappear
in various decompactification limits (or strong coupling limits, in
contexts where the role of the coupling is taken on by one of the
elements 
of $g$). This should lead to new insights regarding the
interconnections between string vacua.

This framework, due to its simplicity and the possibility to treat the
general situation in a unified way, offers good prospects for
nonperturbative extensions of models with spontaneously broken
supersymmetries. It may also have interesting applications of purely
mathematical nature, related to rather nontrivial lattice
identities. These topics, as well as the obvious orbifoldization of
this class of theories, are the subject of current research.

\newpage

\begin{center}
  {\bf Acknowledgements}
\end{center}

We would like to thank Dimitris Matalliotakis and Stephan Stieberger
for helpful discussions. This work was partially supported by
European Union grants SC1-CT92-0789 (H.P.N.) and CIPD-CT94-0034
(M.S.). We thank the CERN Theory Division for hospitality while this
work was completed.

\newcommand{\bi}[1]{\bibitem{#1}}

\newcommand{\plb}[3]{{\em {Phys. Lett.}} {\bf B {#1}} (19{#2}) {#3}.} 
\newcommand{\npb}[3]{{\em {Nucl. Phys.}} {\bf B {#1}} (19{#2}) {#3}.} 
\newcommand{\prl}[3]{{\em Phys. Rev. Lett.} {\bf {#1}} (19{#2}) {#3}.} 
\newcommand{\cmp}[3]{{\em {Comm. Math. Phys.}} {\bf {#1}} (19{#2}) {#3}.} 
\newcommand{\mpla}[3]{{\em Mod. Phys. Lett.} {\bf A {#1}} (19{#2}) {#3}.}
\newcommand{\ijmpa}[3]{{\em Int. J. Mod. Phys.} {\bf A {#1}} (19{#2}) {#3}.}

\newcommand{\prd}{\mbox{\em {Phys. Rev.} {\bf D}}}

\end{document}